\documentclass[fleqn,twoside]{article}
\usepackage{espcrc2}

\usepackage{graphicx}
\usepackage[figuresright]{rotating}
\usepackage{epsfig}
\usepackage{graphicx}
\usepackage{epsf}
\usepackage{amsfonts}
\usepackage{amssymb}
\usepackage{amsbsy}

\newcommand{\AmS}{{\protect\the\textfont2
  A\kern-.1667em\lower.5ex\hbox{M}\kern-.125emS}}
\newcommand{\eq}{\begin{equation}}
\newcommand{\en}{\end{equation}}
\newcommand{\eqa}{\begin{eqnarray}}
\newcommand{\ena}{\end{eqnarray}}
\newcommand{\half}{\frac{1}{2}}
\newcommand{\tr}{\mbox{Tr}}

\newcommand{\bea}{\begin{eqnarray}}
\newcommand{\eea}{\end{eqnarray}}

\title{
\thispagestyle{empty}
\vspace{-25mm}
\rightline{\small DESY 05-246~~~~~}
\rightline{\small ITEP-LAT/2005-25~~~~~}
\vspace{10mm}
New features of the maximal abelian projection
\thanks{Talk given by V.G.B. at the Workshop on Computational Hadron Physics, 
Nicosia, Cyprus.}
}     

\author{V.G.~Bornyakov\address[IHEP]{Institute for High Energy Physics, Protvino 
142281, Russia },
M.I.~Polikarpov\address[ITEP]{Institute of Theoretical and  Experimental 
Physics, Moscow, 117259, Russia},
G.~Schierholz\address{Deutsches Elektronen-Synchrotron DESY, 22603 Hamburg, Germany,\\
John von Neumann-Institut f\"ur Computing NIC/DESY, 15738 Zeuthen, Germany},
T.~Suzuki\address{Institute for Theoretical Physics, Kanazawa University,
Kanazawa 920-1192, Japan},
and
S.N.~Syritsyn\addressmark[ITEP]}

\begin{document}

\begin{abstract}
After fixing the Maximal Abelian gauge in $SU(2)$ lattice gauge theory
we decompose the nonabelian gauge field
into the so called monopole field and the modified nonabelian field with monopoles
removed. We then calculate respective static potentials and find
that the potential due to the modified nonabelian field is nonconfining
while, as is well known, the monopole field potential is linear.
Furthermore, we show that the sum of these potentials approximates the nonabelian 
static potential with 5\% or higher precision at all distances considered. 
We conclude that at large distances the monopole field potential describes 
the classical energy of the hadronic string while the modified nonabelian field 
potential describes the string fluctuations.
Similar decomposition was observed to work for the adjoint static potential.
A check was also made of the center projection in the direct center gauge.
Two static potentials, determined by projected $Z_2$ and by modified nonabelian
field without $Z_2$ component were calculated. It was found that their sum 
is a substantially worse approximation of the $SU(2)$ static potential than 
that found in the monopole case. 
It is further demonstrated that similar decomposition can be made for the flux 
tube action/energy density. 

\vspace{1pc}
\end{abstract}

\maketitle

\section{Introduction}
It is well known \cite{suzuki1,suzuki2,bbms,bm} that after Abelian projection 
in the Maximal Abelian gauge (MAG) \cite{Kronfeld:1987ri,thooft2} the abelian 
string tension, calculated 
from the Abelian static potential, is very close to the nonabelian
string tension and corresponding Coulomb term coefficient is about 1/3 of 
that in the nonabelian static potential. The former observation as many 
others supports the Abelian dominance (for review see e.g. \cite{review}).
It was further discovered \cite{suzuki3,stack,bbms} that the monopole static 
potential also has a string tension close to the nonabelian one and small 
Coulomb term coefficient.  These 
observations are in agreement with conjecture that monopole degrees of 
freedom are responsible for confinement \cite{thooft}. 

The role of the other, nonrelated to monopoles, degrees of freedom was
investigated to a lesser extent so far. We believe that studying the
properties of the observables constructed out of these nonmonopole degrees
of freedom will help to understand the properties of the infrared effective
action.

It is then interesting to see what kind of static
potential one gets if one switches off the monopole contribution to the 
gauge field, i.e. if only off-diagonal gluons and the so called photon part 
of the abelian gluon field are left interacting with static quarks. 

Previously computations of this kind were made in 
\cite{miyamura,Kitahara:1998sj}, where it 
was shown that the topological charge, chiral condensate and
effects of chiral symmetry breaking in quenched light hadron spectrum
disappear after 
removal of the monopole contribution from the relevant operators. Similar 
computations were made within the scope of the $Z_2$ projection studies 
\cite{deforcrand}. It was shown that the modified gauge field with projected 
center vortices (P-vortices) removed produces Wilson loops without area law, 
i.e. lacking 
confinement property. In fact we want to do a similar thing with monopoles. 
But we go one step further -- we consider the sum of the static potentials 
$V_{mod}(R)$ and $V_{mon}(R)$, obtained from the Wilson loops of the 
modified gauge field (with monopole contribution removed) and from the 
Wilson loops of the monopole gauge field alone, respectively. 
We discover that $V_{mod}(R)$ can be well fitted by pure Coulomb
term and the sum $V_{mod}(R)+V_{mon}(R)$ is a good approximation 
of the nonabelian static potential $ V(R)$ at all distances considered in
our measurements. 
Furthermore, we find that a similar approximate decomposition holds for the adjoint 
static potential. We also check the decomposition of the static potential 
induced by the center projection. The approximate decomposition for the static 
potential suggests the possibility of such decomposition for the flux tube 
action/energy density. We also check this possibility.

The paper is organized as follows. In the next section we introduce the 
necessary definitions and describe details of our computations. In section 3 
results for the various static potentials are presented. Section  4 is devoted
to the action/energy density results.  Finally, we conclude in section 5.

\section{Definitions and simulation details}

We study SU(2) lattice gauge theory with Wilson action. The abelian 
projection is made after fixing to the MAG. The abelian projection means coset 
decomposition of the nonabelian lattice gauge field $U(s,\mu)$ into abelian 
field $u(s,\mu)$ and coset field $C(s,\mu)$:
\eq
U(s,\mu) = C(s,\mu) u(s,\mu)\,.
\label{coset}
\en
The Abelian gauge field can be decomposed into a monopole (singular)
part $u_{mon}(s,\mu)$ and photon (regular) part $u_{ph}(s,\mu)$:
\eq
u(s,\mu) = u_{mon}(s,\mu) u_{ph}(s,\mu)
\en
or for the corresponding angles:
\eq
\theta(s,\mu) = \theta_{mon}(s,\mu) + \theta_{ph}(s,\mu)\,,
\en
where $\theta(s,\mu), \theta_{mon,ph}(s,\mu) \in (-\pi,\pi]$ are defined by 
relations
\eqa
u(s,\mu)&=& e^{i \theta(s,\mu)\sigma_3},\\
u_{mon,ph}(s,\mu) &=& e^{i\theta_{mon,ph}(s,\mu)\sigma_3}. 
\ena
$\theta_{mon}(s,\mu)$ satisfies the equation \cite{svs}:
\eq
\partial_{\nu} \partial_{\nu}^{'} \theta_{mon}(s,\mu) -
\partial_{\mu} \partial_{\nu}^{'} \theta_{mon}(s,\nu) =
\label{eqmonf}
\en
\vspace{-5mm}
\[ 2 \pi \partial_{\nu}^{'} m(s,\nu\mu)\,,\]
where $\partial_{\mu}~(\partial_{\mu}^{'})$ are lattice forward (backward)
derivatives. The Dirac plaquette variable $m(s,\mu\nu) \in {\mathbb{Z}}$ is 
determined by decomposition of the abelian plaquette angle 
$ \theta(s,\mu\nu) \equiv \partial_{\mu} \theta(s,\nu)-\partial_{\nu} \theta(s,\mu)$ into regular and singular parts:
\eq
\theta(s,\mu\nu) = \overline{\theta}(s,\mu\nu) + 2\pi m(s,\mu\nu)\,,\\
\label{split}
\en            
\[ \overline{\theta}(s,\mu\nu) \in (-\pi,\pi] \] 
Equation (\ref{eqmonf}) has solution
\eq
\theta_{mon}(s,\mu) = -2 \pi \sum_{s^\prime} D(s-s^\prime) \partial_{\nu}^{'}
m(s^{\prime},\nu\mu) \,,
\label{factor3}
\en
where $D(s)$ is the lattice inverse Laplacian. 
This solution satisfies the Landau gauge condition 
$\partial'_{\mu} \theta_{mon}(s,\mu) = 0$.
The monopole gauge field $\theta_{mon}(s,\mu)$ defined in eq.(\ref{factor3}) 
reproduces more than
95\% of monopoles on our lattices and it reproduces all monopoles
in the infinite volume limit. This explains its name.
We calculate the usual Wilson loops:
\eq
W(C) = \half \tr {\left(\prod_{l \in C} U(l)\right)} \,,
\label{wnonab}
\en
as well as monopole Wilson loops:
\eq
W_{mon}(C) = \half \tr\left(\prod_{l \in C} u_{mon}(l)\right) \,,
\label{wmon}
\en
and the nonabelian Wilson loops with monopole contribution removed:
\eq
W_{mod}(C) = \half \tr  {\left(\prod_{l \in C} \tilde{U}(l)\right)}\,,
  \label{woff}
\en
where the modified nonabelian gauge field is defined as 
\eq
\tilde{U}(s,\mu) = C(s,\mu)\, u_{ph}(s,\mu)\quad .
\label{coset2}
\en
Note that $u_{ph}(s,\mu)$ is abelian projection
of $\tilde{U}(s,\mu)$ and it has very few monopoles in the finite volume and
no monopoles in the infinite volume.

Fixing MAG leaves unbroken $U(1)$ gauge symmetry. 
The general form of this transformation is
\eq
\theta^\prime(s,\mu)  = \theta(s,\mu)  +  \partial_{\mu} \omega(s)  + 2\pi n(s,\mu)\,,
\en
where 
$\theta^\prime(s,\mu), \omega(s) \in (-\pi,\pi]$,  $ n(s,\mu) = 0, \pm 1$. 
Respectively, the Dirac plaquette variable changes as   
\eq
m^\prime(s,\nu,\mu) = m(s,\nu,\mu) + \partial_{\nu} n(s,\mu) -
\partial_{\mu} n(s,\nu) \,. \\
\label{monf2}
\en
Thus the monopole field eq.(\ref{factor3}) depends on the choice of the $U(1)$ 
gauge:
\eq
\theta_{mon}^\prime(s,\mu) = \theta_{mon}(s,\mu) 
+ 2 \pi  n(s,\mu) - \delta(s,\mu) \,, \\
\label{monch}
\en
where 
\eqa
\delta(s,\mu) =&-& 2 \pi \sum_{s^\prime} D(s-s^\prime)
\partial_{\mu} \partial_{\nu}^{'} n(s^\prime,\nu) \\
               &-&\frac{1}{N_{sites}}\sum_s  n(s,\mu)\,.
\ena
The monopole Wilson loop $ W_{mon}(C)$ is invariant under the change of the
monopole field (\ref{monch}) due to the fact that $\sum_{l\in C} \delta(l) =0$.
This is not true for $ W_{mod}(C)$ because of its nonabelian character. 
Indeed, the transformation for $\tilde{U}(s,\mu)$ is
\eq
\tilde{U}^\prime (s,\mu)  = \Omega(s) \tilde{U}(s,\mu) \Omega(s+\mu)^\dagger \Delta(s,\mu)\,,
\en 
where
\eq
\Omega(s) = {\rm diag} \{e^{(-i\omega(s))}, e^{(i\omega(s))} \} \,,
\en
\eq
\Delta(s,\mu) = {\rm diag} \{e^{-i\delta(s,\mu)}, e^{i\delta(s,\mu)}\} \,.
\en
This transformation is not a gauge transformation and thus the Wilson loop 
$W_{mod}$ depends on the choice of the gauge used in the definition of the 
monopole field eq.(\ref{factor3}). This problem is solved by using  
in (\ref{factor3}) the Dirac plaquettes $m(s,\mu\nu)$ in some particular gauge. 
We choose the Landau gauge defined by the gauge condition:
\eq
\max_{\omega} \sum_{s,\mu} cos(\theta^\prime(s,\mu)) \,.
\label{landau}
\en
Another possible gauge condition would be the minimization of the number of the Dirac plaquettes:
\eq
\min_{n_\mu} \sum_{s,\mu,\nu} (m^\prime (s,\mu,\nu))^2 \,.
\en
Up to Gribov copies both conditions fix configuration of Dirac plaquettes 
$m(s,\mu\nu)$ completely and thus fix $\theta_{mon}(s,\mu)$. 
In general, results for $W_{mod}$ can be different for different gauge 
conditions 
which fix the Dirac plaquette configuration. But for two gauge conditions 
introduced above we may hope that results  are similar because fixing $U(1)$ 
Landau gauge also strongly reduces the number
of Dirac plaquettes and thus we  expect that for given lattice configuration 
number of Dirac plaquettes in these two gauges are close to each other.
 
We calculated $R\times T$ rectangular Wilson loops $W(R,T)$, $W_{mon}(R,T)$ 
and $W_{mod}(R,T)$.
To extract the nonabelian static potential $V(R)$ link integration 
\cite{parisi} and smearing \cite{ape}
have been employed. Smearing was used in computations of the 
modified field potential $V_{mod}(R)$. 
Computations were done at $\beta=2.5$ on $24^4$ lattices using 100 statistically 
independent configurations.  To fix MAG 10 randomly generated gauge 
copies fixed by simulating annealing algorithm \cite{bbms} were used. 

\section{Static potential decomposition}
\begin{figure*}[htbp]
\begin{center}
\hspace{-0.8cm}\epsfxsize=8.5truecm \epsfysize=8.truecm \epsfbox{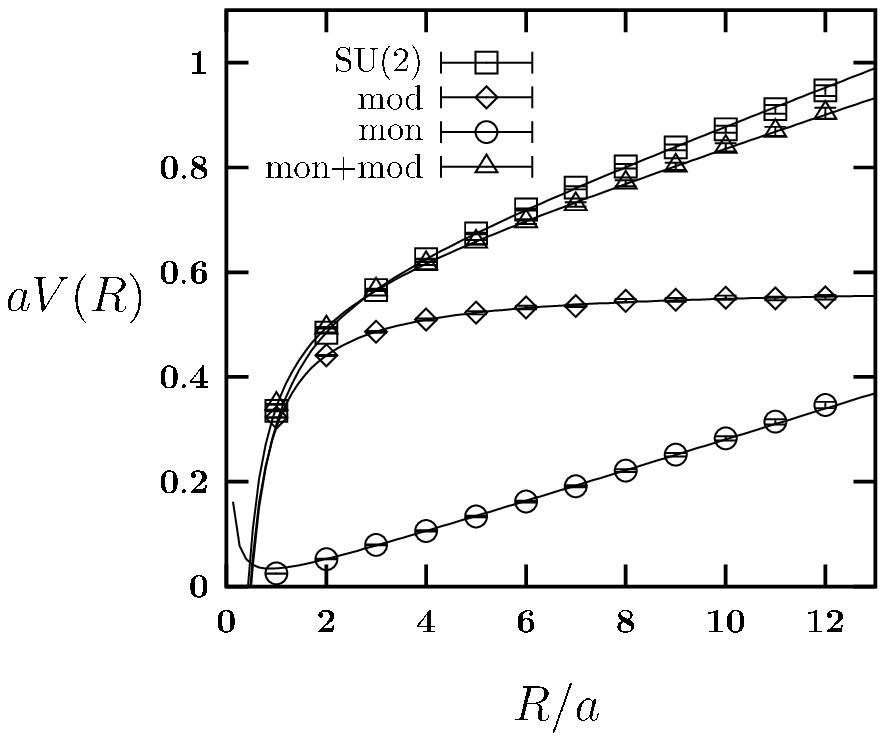}
\hspace{-0.5cm}\epsfxsize=8.5truecm \epsfysize=8.truecm \epsfbox{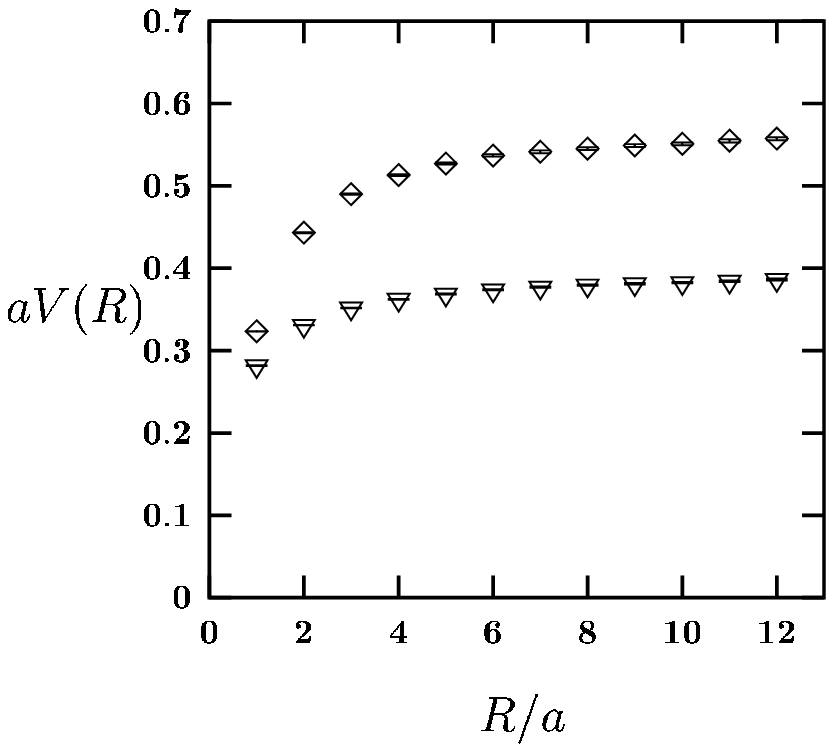}
\vspace{-1.5cm}
\caption{(left) Comparison of the nonabelian potential $V(R)$ (squares) and the 
sum  $V_{mod}(R)+V_{mon}(R)$ (triangles). 
$V_{mod}(R)$ (diamonds) and $V_{mon}(R)$ (circles) are also depicted.\,\,
(right) $V_{mod}$ after U(1) gauge fixing
(diamonds) and without  U(1) gauge fixing (triangles).
}
\label{figure1}
\end{center}
\end{figure*}

In Fig.~\ref{figure1}(left) we show the monopole $V_{mon}(R)$ and the modified 
field $V_{mod}(R)$ potentials.
We find that $V_{mon}(R)$ is linear at large distances and has small 
curvature at small distances, as was observed many times before. Our result 
for  $V_{mod}(R)$ is the first result for this potential. It can be seen 
from Fig.~\ref{figure1}(left) that this potential is of the Coulombic type. 
Indeed it can be very well fitted by 
$V^{mod}_0 - \alpha_{mod}/R$ with $\alpha_{mod}=0.274(9)$.
The fitting curve is shown in Fig.~\ref{figure1}(left). Thus, removing the 
monopole contribution from the Wilson loop operator leaves us with
Wilson loop which has no area law behavior, i.e. confinement property is lost. 
This result is similar to that obtained
in \cite{deforcrand} after removing P-vortices.    

In Fig.~\ref{figure1}(left) we also compare the sum $ V_{mon}(R)+V_{mod}(R)$ 
with the nonabelian potential $V(R)$.  One can see that the nonabelian 
static potential is well approximated by this sum, i.e.
\eq
V(R) \approx V_{mon}(R)+V_{mod}(R)\,.
\label{decomp}
\en
This is our main result.
On the quantitative level, we find that 
\eq
\frac{|\delta V(R)|}{V(R)} < 0.05\,,
\en
where $\delta V(R) = V(R) - \left( V_{mon}(R)+V_{mod}(R) \right)$.
This observation can be formulated in the following way:
the potential for the static  sources, interacting with  the nonabelian gauge 
field $U(s,\mu)$ can be approximated
by the sum of the potential for the sources, interacting only with the 
monopole field $u_{mon}(s,\mu)$, and potential for the sources, 
interacting only with the modified field $\tilde{U}(s,\mu)$. 
All calculated potentials were fitted by the usual linear plus Coulomb functions.
Results for the fit parameters are presented in Table~\ref{results}.

\begin{table*}[tbph]
\caption{Parameters of the potentials obtained by fits with function
$V_0-\alpha/R+\sigma R$.} 
\vspace{0.5cm}

\begin{tabular}{clll}
\hline
                   & ~~~~~$\sigma a^2$  & ~~~~$\alpha$ & ~~~$aV_0$ \\
\hline
$aV(R)$             & ~0.0339(4) & ~0.286(4)  & ~0.560(2) \\ 
$aV_{mon}(R)+aV_{mod}(R)$  & ~0.0310(3)& ~0.264(6) & ~0.577(4) \\
$aV_{mon}(R)$          & ~0.0311(3) & -0.018(4) & -0.006(3) \\
$aV_{mod}(R)$          & -0.0002(2)& ~0.280(10) & ~0.583(5) \\
$aV_{Z_2}(R)+aV_{mod,Z_2}(R)$& ~0.0224(15)& ~0.334(20)& ~0.668(15) \\
$aV_{Z_2}(R)$           & ~0.0249(6) & -0.024(7) & -0.003(4) \\
$aV_{mod,Z_2}(R)$       & -0.029(6) & ~0.383(10) & ~0.686(6)  \\
\hline
\end{tabular}
\label{results}
\end{table*}

In Fig.~\ref{figure1}(right) we compare $V_{mod}(R)$ calculated with and without 
additional U(1) gauge fixing defined by eq.~(\ref{landau}). One can see that without U(1) gauge fixing
$V_{mod}(R)$ is substantially lower than it is after  U(1) gauge fixing. Fitting  with constant plus
Coulomb fitting function we found the Coulomb coefficient two times smaller.   
Thus the approximate  decomposition of the potential (\ref{decomp})
does not hold without this gauge fixing.

The approximate relation (\ref{decomp}) implies that for large $T$:
\eq
\langle W(R,T) \rangle \approx \kappa~\langle W_{mon}(R,T) \rangle~ 
\langle W_{mod}(R,T) \rangle 
\label{wl1}
\en
Indeed we found that for $R,T \ge 3$ our data for unsmeared 
$\langle W(R,T) \rangle$ can be 
fitted by the right hand side of eq.~(\ref{wl1}) (with unsmeared $W_{mon}$ and 
$W_{mod}$) with $\kappa=0.886(9)$.

\begin{figure*}[htbp]
\begin{center}
\hspace{-0.9cm}\epsfxsize=8.5truecm \epsfysize=8.truecm \epsfbox{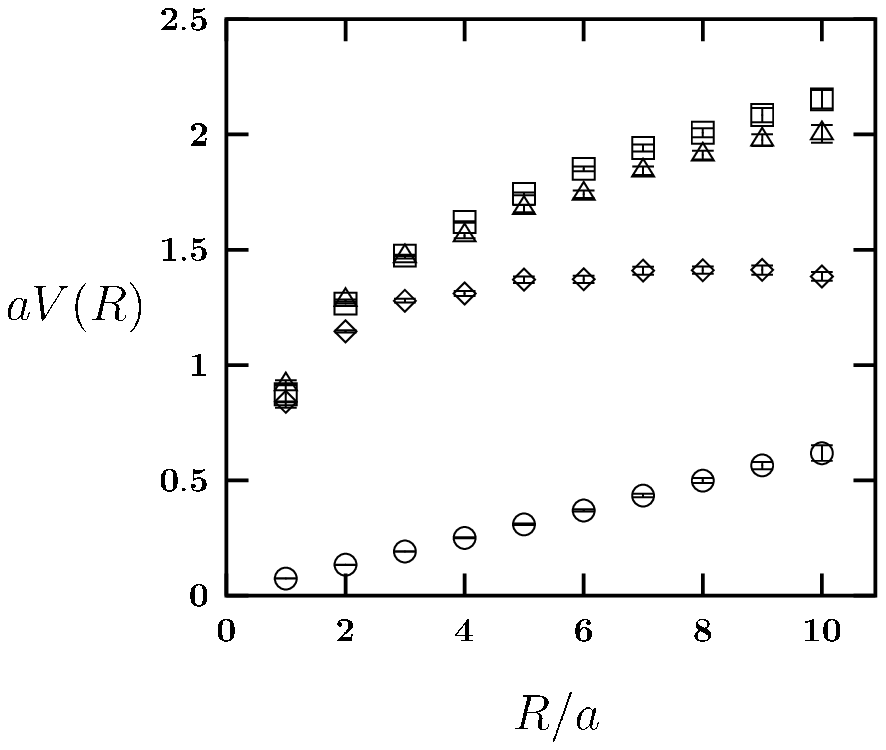}
\hspace{-0.4cm}\epsfxsize=8.5truecm \epsfysize=7.9truecm \epsfbox{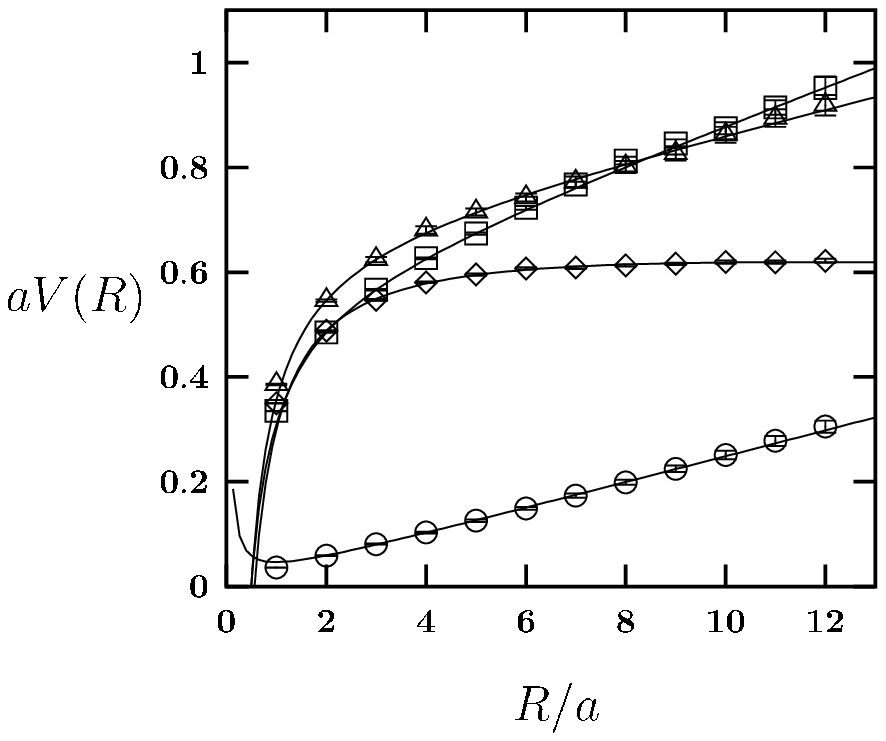}
\vspace{-1.5cm}
\caption{(left) The adjoint nonabelian potential $V_{adj}(R)$ 
(squares) is compared with the
sum  $V_{mod,adj}(R)+V_{mon,2}(R)$ (triangles).
$V_{mod,adj}(R)$ (diamonds) and $V_{mon,2}(R)$ (circles) are also 
shown. (right) Comparison of the nonabelian potential $V(R)$ (squares) and the
sum  $V_{mod,Z_2}(R)+V_{Z_2}(R)$ (triangles).
Also depicted are  $V_{mod,Z_2}(R)$ (diamonds) and $V_{Z_2}(R)$ (circles).}
\label{figure2}
\end{center}
\end{figure*}

Next we want to address the adjoint potential decomposition. 
We calculated nonabelian adjoint Wilson loop $W_{adj}(R,T)$, 
charge two monopole Wilson loop
\eq
W_{mon,2}(R,T) = W_{mon}^2(R,T) 
\label{wmonadj}
\en
and adjoint Wilson loop $W_{mod,adj}(R,T)$ for the modified nonabelian 
field $\tilde{U}(s,\mu)$. From these Wilson loops the respective potentials
$V_{adj}(R)$,  $V_{mon,2}(R)$ and $V_{mod,adj}(R)$ were extracted. 
In Fig.~\ref{figure2}(left) we compare the sum $V_{mon,2}(R) + 
V_{mod,adj}(R)$ with the adjoint potential $V_{adj}(R)$. As in the 
fundamental case we see approximate decomposition:
\eq
 V_{adj}(R) \approx V_{mon,2}(R) + V_{mod,adj}(R)\,.
\label{decomp2}
\en
The potential $ V_{mod,adj}(R)$ looks purely Coulombic. A Coulomb term
coefficient $\alpha_{mod,adj}=0.69(4)$ was found by fitting the data at 
$R \ge 2$. $ V_{mon,2}(R) $ is
linear with small curvature as was observed before in \cite{bbms}.
Our result (\ref{decomp2}) supports the conjecture \cite{poulis,bbms} that 
the abelian charge
two potential should be considered as the abelian projection for the adjoint 
potential.

As we noted above, in Ref. \cite{deforcrand} it was shown that 
in the central gauge after removal of 
P-vortices the Wilson loop looses confinement property. In the MAG we found 
that removal of monopoles also leaves the Wilson
loop without confinement property. Moreover, we found that the approximate 
decomposition
(\ref{decomp}) holds. It is then interesting to check similar decomposition for
the central gauge. We made  computations in the Direct Central (DC) gauge 
at $\beta=2.5$ using half of the set of configurations which were used for 
computations in the MAG. Results are presented in Fig.~\ref{figure2}(right).
From comparison of Fig.~\ref{figure1}(left) and Fig.~\ref{figure2}(right) and from
comparison of the 
respective fitting parameters, presented in Table~\ref{results}, 
one can see that approximate decomposition works substantially better
in MAG. This is not unexpected for large distances, since it is 
known that in the center gauges the P-vortex string tension is smaller 
than the monopole string tension in MAG.

Note that decomposition similar to eq.(\ref{decomp2}) is impossible for central
projection since the charge two central projected Wilson loop is identity.

\section{Flux tube profile}
In this section we present our results for the flux tube action/energy density decomposition.

\begin{figure*}[htbp]
\begin{center}
\hspace{-0.9cm}\epsfxsize=8.3truecm \epsfysize=7.3truecm \epsfbox{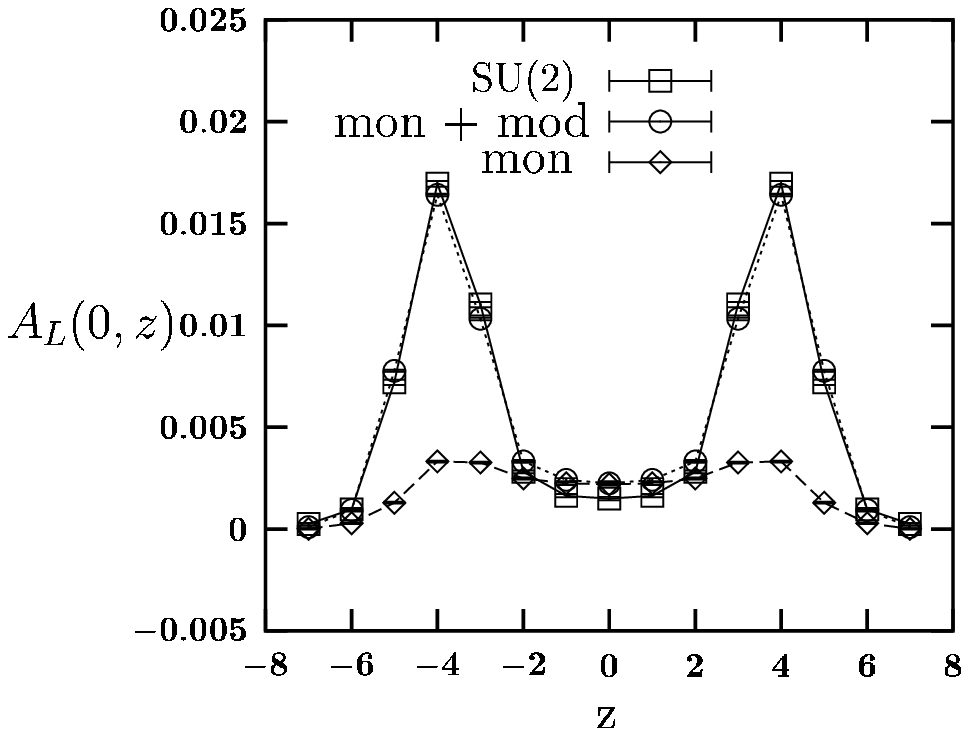}
\hspace{-0.3cm}\epsfxsize=8.3truecm \epsfysize=7.3truecm \epsfbox{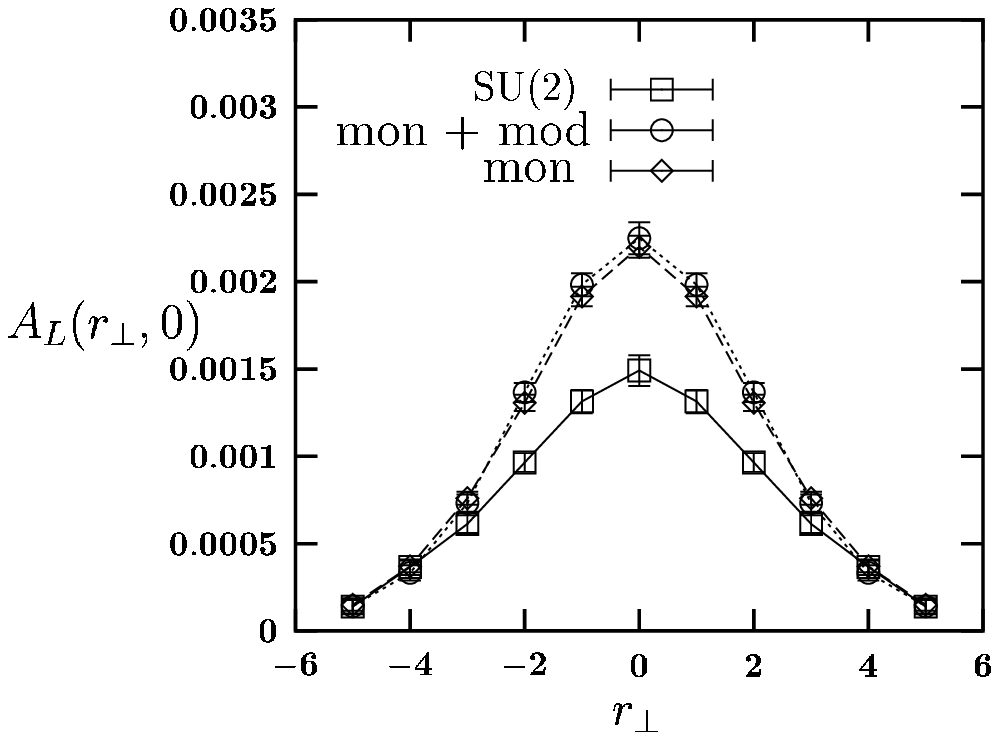}
\vspace{-2.0cm}
\caption{The longitudinal action densities: nonabelian (squares), monopole 
(diamonds), and sum of the monopole and modified (circles). (left): as functions of
the coordinate $z$ along the $Q \bar{Q}$ axes; (right): as functions of the 
distance $r_\perp$ from this axis.}
\label{figure3}
\end{center}
\vspace{-1cm}
\end{figure*}

\begin{figure*}[htbp]
\begin{center}
\hspace{-0.7cm}\epsfxsize=8.5truecm \epsfysize=7.5truecm \epsfbox{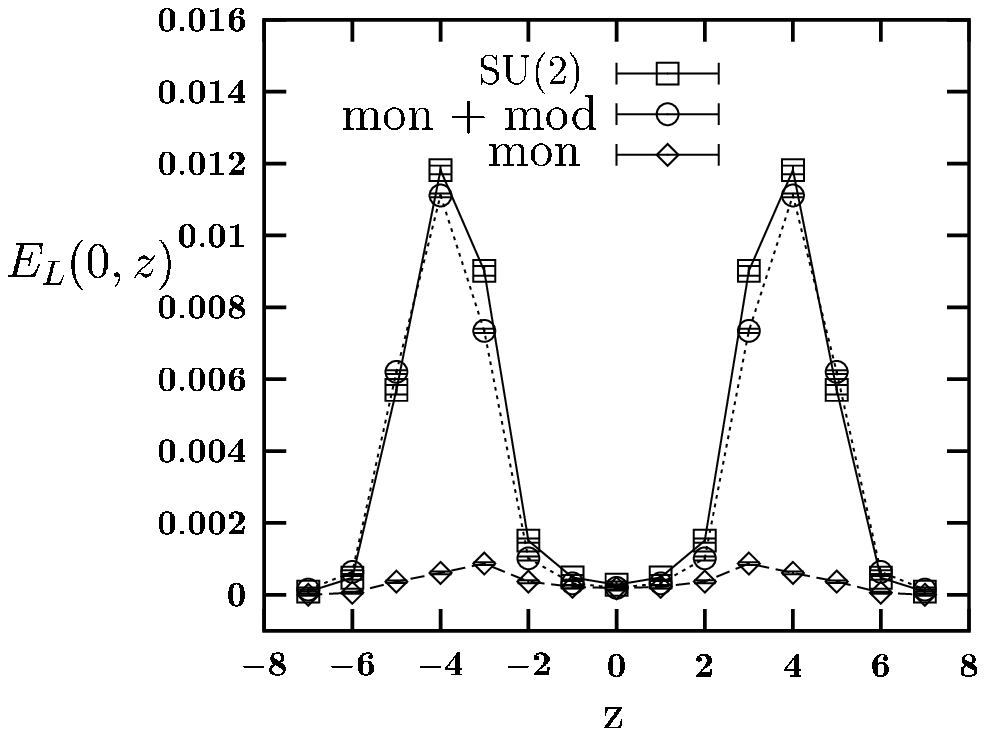}
\hspace{-0.4cm}\epsfxsize=8.5truecm \epsfysize=7.5truecm \epsfbox{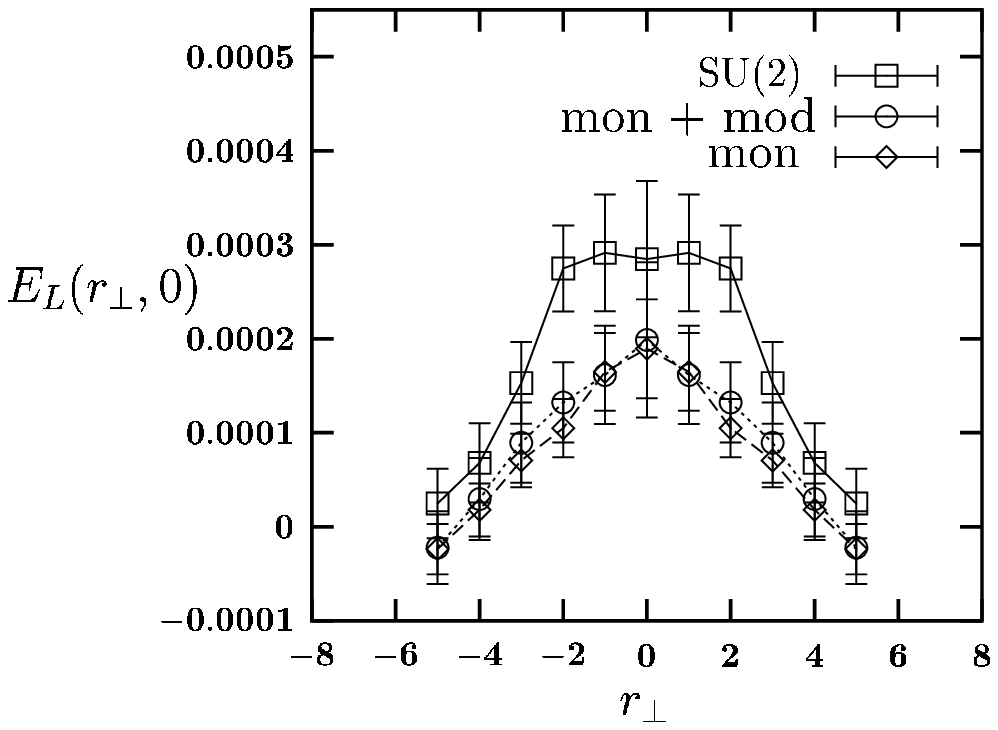}
\vspace{-2.0cm}
\caption{Same as in Fig.~\ref{figure3} but for the longitudinal energy densities.}
\label{figure4}
\end{center}
\end{figure*}

Taking derivative with respect to $\beta$ on the left and right hand sides of 
eq. (\ref{wl1}) and dividing by respective expressions we obtain 
(for $T \to \infty$) \footnote{In this derivation we take into account that the 
Faddeev-Popov determinant has no explicit dependence on $\beta$}:
\[ \frac{\langle S \, W(R,T) \rangle}{ W(R,T) } - \langle S \rangle  \approx 
\left( \frac{\langle S \, W_{mon}(R,T) \rangle}{ W_{mon}(R,T) } - \langle S \rangle \right)\times \]
\eq
\left( \frac{\langle S \, W_{mod}(R,T) \rangle}{ W_{mod}(R,T) } - \langle S \rangle \right)\,,
\en
where $S$ is the nonabelian action averaged over the lattice.  It is then 
interesting to check whether such approximate decomposition is  
valid also locally, i.e. whether the following approximate relation holds:
\[\langle F^2(\vec{s},\mu,\nu) \rangle _W \approx  \]
\eq
\langle F^2_{mon} (\vec{s},\mu,\nu) \rangle _W+
\langle F^2_{mod} (\vec{s},\mu,\nu) \rangle _W \,,
\label{f2} 
\en
where
\[ \langle F^2_f(\vec{s},\mu,\nu) \rangle _W \equiv
\frac{\langle W_f \, F^2 (s,\mu,\nu) \rangle }{\langle W_f \rangle} - 
\langle F^2 (s,\mu,\nu) \rangle\,,  \]
the lattice field strength squared is defined as
\eq
\frac{1}{2} F^2(s,\mu,\nu) = 1 - \half \tr~U(s,\mu,\nu)\,, 
\en
$ F^2_{mon}(s,\mu,\nu) $ and $ F^2_{mod}(s,\mu,\nu) $ are defined analogously with 
respective plaquettes instead of $ U(s,\mu,\nu) $.
In Fig.~\ref{figure3} and Fig.~\ref{figure4} we show various longitudinal action 
and energy densities defined as follows:
\[ A_L(r_\perp,z) = \frac{1}{2}\left( \langle F^2 (\vec{s},0,3) \rangle _W  + 
\langle F^2 (\vec{s},1,2) \rangle _W \right) \,, \]
\[ E_L(r_\perp,z) = \frac{1}{2}\left( \langle F^2 (\vec{s},0,3) \rangle _W  - 
\langle F^2 (\vec{s},1,2) \rangle _W \right) \,, \]
where $r_\perp$ is the distance from the quark-anti-quark axes, $z$ is 
coordinate along this axes. The distance between sources is $R/a=8$. 
One can see from these figures that relation (\ref{f2})
indeed holds with rather good precision. Similar results were obtained for the
transverse action and energy densities. In the computations of the
action/energy density Wilson loops with $T/a=4$ were used for the nonabelian and
modified cases and with $T/a=6$ for the monopole case.


\section{Conclusions and discussion}

Using abelian projection after fixing MAG we shown that the static 
potential $V(R)$ can be approximately decomposed into two components, 
see eq.~(\ref{decomp}). Our preliminary results at smaller lattice spacing 
($\beta=2.6$) indicate that in the continuum limit the agreement between
left and right hand sides in (\ref{decomp}) improves.
One term in (\ref{decomp}), $V_{mon}(R)$, is due to the monopole gauge 
field (\ref{factor3}) contribution,
another one, $V_{mod}(R)$, is due to the contribution of the modified
 nonabelian gauge field $\tilde{U}(s,\mu)$ with monopoles removed. 
Comparison of the parameters 
of these potentials, given in Table \ref{results}, with the effective string 
model predictions suggests 
the following interpretation: at large distances $V_{mon}(R)$ is the classical
energy of the flux tube and $V_{mod}(R)$ is the flux tube fluctuations energy.
This conclusion is supported by our results for the adjoint static potential,
see Fig.~\ref{figure2}(left).

It was checked whether a similar decomposition of $V(R)$  holds for the center 
projection after fixing to DC gauge. We calculated projected $Z_2$ potential 
$V_{Z_2}(R)$ and modified nonabelian field potential $V_{mod,Z_2}(R)$. 
The latter was previously introduced and 
measured in \cite{deforcrand}. We found that decomposition in this case
holds with substantially less precision than in the monopole case. 
This can be seen from
comparison of Figs.~\ref{figure1}(left) and \ref{figure2}(right), and from 
Table \ref{results}. Both the string tension and the Coulomb coefficient
obtained from the fits of the sum  $V_{Z_2}(R)+V_{mod,Z_2}(R)$ are in worse 
agreement with respective parameters of the nonabelian potential than those 
obtained from the fits of the sum $V_{mon}(R)+V_{mod}(R)$.

The decompositions eq.~(\ref{decomp}) and eq.~(\ref{decomp2}) are similar
to decomposition in the compact $U(1)$ model \cite{Banks:1977cc} into the 
photon and monopole components of the static potential. In that model the
action can be respectively decomposed into the monopole and photon parts
without interaction term. 
Contrary to the compact $U(1)$
theory such term is unavoidable in the infrared effective action of 
the $SU(2)$ gluodynamics since the charge two component
of the off-diagonal field $C(s,\mu)$ provides the string breaking of
the charge two monopole potential $V_{mon,2}$ \cite{Chernodub:2003sy}.
The relations eq.~(\ref{decomp}) and 
eq.~(\ref{decomp2}) are hinting that this interaction term should be weak.

Our results for the action and energy densities in the quark-antiquark system,
presented in Fig.~\ref{figure3} and Fig.~\ref{figure4} 
show that the approximate decomposition holds also for these quantities.

\vspace{1.0cm}
\noindent
{\bf\large  Acknowledgments}\\
The present work was  supported by grants  
RFBR 03-02-16941, RFBR 04-02-16079, RFBR 05-02-16306a, RFBR 05-02-17642, 
RFBR-DFG 03-02-04016, DFG-RFBR 436RUS113/739/0,
and by the EU Integrated Infrastructure Initiative Hadron Physics (I3HP).

\end{document}